\documentclass[prb,preprint,groupedaddress]{revtex4}
\usepackage{amsmath} 
\usepackage{amssymb} 
\usepackage{amsfonts}
\usepackage[dvips]{graphicx} 
\usepackage[]{epsfig} 
\begin{document}

\title{Coupling nonpolar and polar solvation free energies in implicit solvent models}
 \author{J. Dzubiella} \email[e-mail address:]
{jdzubiella@ucsd.edu} \affiliation{NSF Center for Theoretical
Biological Physics (CTBP),} \affiliation{Department of Chemistry and
Biochemistry, University of California, San Diego, La Jolla,
California 92093-0365} \author{J.~M.~J. Swanson} \affiliation{NSF Center for
Theoretical Biological Physics (CTBP),} \affiliation{Department of
Chemistry and Biochemistry, University of California, San Diego, La
Jolla, California 92093-0365} \author{J.~A. McCammon} \affiliation{NSF
Center for Theoretical Biological Physics (CTBP),}
\affiliation{Department of Chemistry and Biochemistry, University of
California, San Diego, La Jolla, California 92093-0365}

\date{\today}

\begin{abstract}
Recent studies on the solvation of atomistic and nanoscale solutes
indicate that a strong coupling exists between the hydrophobic,
dispersion, and electrostatic contributions to the solvation free
energy, a facet not considered in current implicit solvent models. We
suggest a theoretical formalism which accounts for coupling by
minimizing the Gibbs free energy of the solvent with respect to a
solvent volume exclusion function. The resulting differential equation
is similar to the Laplace-Young equation for the geometrical
description of capillary interfaces, but is extended to microscopic
scales by explicitly considering curvature corrections as well as
dispersion and electrostatic contributions.  Unlike existing implicit
solvent approaches, the solvent accessible surface is an output of our
model. The presented formalism is illustrated on spherically or
cylindrically symmetrical systems of neutral or charged solutes on
different length scales.  The results are in agreement with
computer simulations and, most importantly, demonstrate that our
method captures the strong sensitivity of solvent expulsion and
dewetting to the particular form of the solvent-solute interactions.
\end{abstract}
\maketitle
\section{Introduction}
Implicit solvent models are widely used in theoretical chemistry to
study the solvation of biomolecular systems, as well described in the
review of Roux and Simonson\cite{roux:biochem}. They provide a more
efficient, although generally less accurate, alternative to
atomistically-resolved explicit solvent simulations.  The solvation
free energy in these models is usually split into nonpolar (np) and
polar (p) terms,
\begin{eqnarray}
\Delta G=\Delta G_{\rm np}+\Delta G_{\rm p},
\end{eqnarray}
which are treated in separate energetic evaluations. The nonpolar term
includes   the   energetic   cost   of   cavity   formation,   solvent
rearrangement,  and solute-solvent  dispersion interactions introduced
when  the uncharged  solute is  brought from  vacuum into  the solvent
environment. The polar term describes  the free energy of charging the
mono- or multipolar solute in the dielectric medium.

The nonpolar term is commonly approximated by surface area models,
i.e.~$\Delta G_{\rm np}\simeq\gamma S$, where $S$ is the
solvent accessible surface area\cite{richards} and $\gamma$ is an
energy per surface area constant, which is a priori not known but
fit to atomistic simulations.  The deficiencies of this simple surface
area approach have been recognized and a further decomposition of the
nonpolar term into cavity (cav) and van der Waals dispersion (vdW) terms has been
proposed,\cite{gallicchio:jpcb,gallicchio:jcc} $\Delta G_{\rm
np}=\Delta G_{\rm cav}+\Delta G_{\rm vdW}$. This approach has shown
improved results for the solvation of
alkanes,\cite{gallicchio:jpcb,zacharias} the alanine
peptide,\cite{su:biochem} and nonpolar native and misfolded
proteins.\cite{levy:jacs} The electrostatic (polar) contribution of
the solvation free energy is often approximated by generalized
Born\cite{bashford} (GB) or Poisson-Boltzmann\cite{sharp:honig} (PB)
models. Both methods use a position-dependent dielectric
constant,\cite{roux:biochem} assigned on the basis of
the solute surface, which  can be defined in several ways,\cite{baker}
or defined implicitly by integration methods.
It has been emphasized that all three contributions, $\Delta G_{\rm
cav}$, $\Delta G_{\rm vdW}$, and $\Delta G_{\rm p}$ depend critically
on the location of the solvent-solute interface.  It has {\it also}
been shown that the effective location of the solvent-solute interface
can vary according to the local electrostatic\cite{nina}
and dispersion\cite{huang:jpcb:2002} potentials. This suggests
that interfacial, dispersion and electrostatic contributions should be
coupled in implicit solvent approaches. The importance of capturing
the right balance between nonpolar and electrostatic contributions
in implicit solvation models was emphasized by Ashbaugh
{\it et al.} in their study of amphiphiles.\cite{ashbaugh:biophys}

The significance of nonpolar and polar coupling becomes even
more evident when solvation is studied on length scales which are
large compared to the solvent molecule (typically $\gtrsim$ 1 nm for
water), where solvent dewetting ('drying') can occur. In this
mechanism, first envisioned by Stillinger,\cite{stilinger} the solvent
molecules tend to move away from the surface of a large nonpolar
solute forming a liquid-gas like interface parallel to the solute
interface. When the surfaces of two
large solutes come together dewetting can be amplified due to the
gain of interfacial free energy (by decreasing the total
liquid-vapor interface area) giving rise to a strong effective
attraction.\cite{lum:jpc,hummer:jcp:1997,chandler:review} Early evidence of 
confinement-induced dewetting was given only by explicit water
simulations for smooth plate-like solutes with a purely repulsive
solute-solvent interaction.\cite{wallquist:jpc} More recently,
however, it has been demonstrated in varying degrees in several systems with attractive
solute-solvent interactions including smooth parallel plate-like
solutes,\cite{huang:pnas} atomistically resolved
paraffin-plates,\cite{huang:jpcb} graphite-plates\cite{pettitt}, carbon
nanotubes,\cite{hummer:nature} and hydrophobic ion channels.\cite{allen:prl,beckstein:jpc,sukharev}

Several of these studies indicated that the magnitude of dewetting is
sensitive to the nature of the solute-solvent attractive dispersion
interactions.  \cite{huang:pnas,huang:jpcb,pettitt} A similar
sensitivity was found in systems where the solutes carry charges or
are exposed to an external electric field, e.g. electrostatic interactions have
been shown to strongly affect the dewetting behavior of hydrophobic
channels\cite{vaitheesvaran, dzubiella:channel1,dzubiella:channel2}
and hydrophobic spherical nanosolutes.\cite{dzubiella:jcp:2003,
dzubiella:jcp:2004} Furthermore, two recent simulations of proteins
supported the importance of solvent dewetting and its sensitivity in
realistic biomolecular systems. First, a simulation of the two-domain
BphC enzyme showed that the region between the two domains was
completely dewetted when vdW and electrostatic interactions were
turned off, but accommodated 30$\%$ of the bulk density with the
addition of vdW attraction (water was found mainly at the edges of the
considered volume, while the central region was still empty), and
85-90$\%$ with the addition of electrostatic
interactions.\cite{zhou:science} Second, Liu {\it et al.}  observed a
clear dewetting transition in the simulation of the collapse of the
melittin tetramer, which was strongly sensitive to the type and
location of the hydrophobic residues around the dewetted
region.\cite{berne:nature}
 
Considering the aforementioned studies, we postulate that coupling of
the nonpolar and polar solvation contributions in implicit solvent
models is crucial for an accurate determination of solvation free
energies without too many system-dependent fit parameters. We suggest
a general theoretical formalism in which the particular energetic
contributions are coupled.  Similar to the approach of Parker {\it et
al.}  in their study of bubble formation at hydrophobic
surfaces,\cite{attard} we express the Gibbs free energy as a {\it
functional} of the solvent volume exclusion function,\cite{beglov:jcp}
and obtain the optimal solute surface via minimization.  As
we will show, this minimization leads to an expression which is
similar to the Laplace-Young equation for the description of
macroscopic capillary interfaces,\cite{kralchevsky} but is generalized
to explicitly include curvature corrections and solvent-solute
interactions, i.e.  short-range repulsion (excluded volume),
dispersion, and electrostatics.  This extension of the Laplace-Young
theory allows a geometric description of solvation on mesoscopic and
microscopic scales. Related approaches in other fields are the
Helfrich description of vesicle and membrane
surfaces,\cite{helfrich,helfrich2} wetting in colloids and granular
media,\cite{kralchevsky,bieker} and functional treatments of
electrowetting.\cite{electrowetting}

While most implicit solvent approaches define the solute surface with
a geometrical evaluation of the molecular surface, vdW surface, or
canonical solvent accessible surface (SAS),\cite{richards,baker} it is
an output of our theory.  The surface obtained by minimizing our free
energy functional will, in general, be very different than the
aforementioned established surface definitions. In particular, our
solvent accessible surface should not be confused with the canonical
SAS,\cite{richards} which is simply the envelope surrounding
probe-inflated spheres. Similarly, phenomenological continuum theories
applied to solvent dewetting always assume a certain, simplified
geometry for the dry region, e.g. a cylindrical volume for system like
hydrophobic ion channels,
\cite{allen:jcp,beckstein:jpc,dzubiella:channel2} plate-like
particles,\cite{lum:jpc,huang:pnas} or two hydrophobic spherical
solutes.\cite{attard2} For a few simple systems this might be a valid
approximation but for more complicated solute geometries the shape of
the dewetted volume is unknown and a different approach, as suggested
in this work, is necessary. We expect our formalism to be particularly
useful in solvation studies of large protein assemblies where the
hydrophobic surfaces are highly irregular and laced with hydrophilic
units,\cite{gerstein,rossky:nature} and for which a unified
description of hydration on different length scales is
important.\cite{chandler:review} Another potential application is the
solvation of superhydrophobic nanosolutes\cite{super} and
wetting/dewetting in near-critical colloidal mixtures.\cite{bieker}

A brief summary of our work has been published elsewhere.\cite{prl}
Here we present more challenging test cases and an expanded discussion
of the approximations and limitations of this model.  The rest of the
paper is organized as follows: In section II we present our
theoretical formalism and chosen approximations. In section III we
first verify that our method can describe solvation on molecular
scales with noble gases, ions, and small alkanes. We then demonstrate
that it captures the strong sensitivity of dewetting and hydrophobic
hydration to specific solute-solvent interactions on larger scales
with two alkane-assembled spheres. In section IV we conclude with some
final remarks.

\section{Theory}

\subsection{Basic formalism}

Let us consider an assembly of solutes
with arbitrary shape and composition surrounded by a dielectric
solvent in a volume $\cal W$.  Furthermore, we define a subvolume (or
cavity) $\cal V$ empty of solvent for which we can assign a volume
exclusion function given by
\begin{eqnarray}
 v(\vec r)= \begin{cases} 0 & {\text {for}} \,\,\vec r \in \cal V; \cr 1 & {\rm else}. \cr
 \end{cases}
\end{eqnarray}
We assume that the surface surrounding the volume is continuous and
closed, i.e.~has no boundary.  The absolute volume $V$ and surface area $S$
of $\cal V$ can then be expressed as functionals of $v(\vec r)$ via
\begin{eqnarray}
V[v]=\int_{\cal W}{\rm d}^3r \;[1-v(\vec r)] \nonumber \\
S[v]=\int_{\cal W}{\rm d}^3r \;|\nabla v(\vec r)|,
\end{eqnarray}
where $\nabla\equiv\nabla_{\vec r}$ is the usual gradient operator
with respect to the position vector $\vec r$ and $|\nabla v(\vec r)|$
gives a $\delta$-function-like contribution only at the volume
boundary. The expression ${\rm d}^3r \;|\nabla v(\vec r)|\equiv{\rm
d}S$ can thus be identified as the infinitesimal surface element. In
this continuum solvent model the solvent density distribution
is simply $\rho(\vec r)=\rho_0 v(\vec r)$, where $\rho_0$ is the
bulk density of the solvent at the desired temperature and
pressure. Local inhomogeneities of the solvent density, apart from the
zero to $\rho_0$ transition at the volume boundaries, are
neglected. The solutes' positions and conformations are fixed such that
the solutes can be considered as an external potential to the solvent
without any degrees of freedom.

As motivated before, we suggest expressing the Gibbs free energy
$G[v]$ as a functional of the volume exclusion function
$v(\vec r)$, and obtaining the optimal solute volume via
minimization
\begin{eqnarray}
\delta G[v]/\delta v(\vec r)=0,
\label{min}
\end{eqnarray}
where $\delta../\delta v$ denotes the functional derivative with respect to the function $v$. We adopt following
ansatz for the Gibbs free energy of the solvent:
\begin{eqnarray}
\label{eq:grand}
&G[v]&=G_{\rm pr}[v]+ G_{\rm int}[v]+ G_{\rm ne}[v]+G_{\rm es}[v]\\
&=& P V[v]+\int_{\cal W}{\rm d}^3r\;\gamma(\vec r;[v])|\nabla v(\vec r)|\nonumber \\
&+&\rho_0\int_{\cal W}{\rm d}^3r\;v(\vec r) U(\vec r)\nonumber \\ 
&+&\!\!\!\!\!\!\int_{\cal W}\!\!\!{\rm d}^3r\!\left\{\frac{\epsilon_0}{2}\epsilon(\vec r;[v])[\nabla\Psi(\vec r)]^2\!-\lambda(\vec r)\Psi(\vec r)\!+v(\vec r)U_{\rm mi}(\vec r)\right\}\nonumber
\end{eqnarray}
Let us discuss each of the terms in Eq.~(\ref{eq:grand}) in turn.  The
first term, $G_{\rm pr}[v]$, proportional to the volume $V$, is the
energy of creating a cavity in the solvent against the difference in
bulk {\it pressure} between the liquid and vapor phase, $P=P_l-P_v$.
For water in ambient conditions, which is close to the liquid-vapor
transition, this term is relatively small and can generally be
neglected for solutes on molecular scales. The second term $G_{\rm
int}[v]$ describes the energetic cost due to solvent rearrangement
around the cavity {\it interface} with area $S$ in terms of a free
energy/surface area functional $\gamma(\vec r;[v])$. This interfacial
energy penalty is thought to be the main driving force behind
hydrophobic phenomena.\cite{chandler:review} $\gamma$ is a solvent
specific quantity that also depends on the particular topology of the
cavity-solvent interface, i.e.~it varies locally in space and is a
functional of the volume exclusion function $\gamma=\gamma(\vec
r;[v])$.\cite{zwanzig} The exact form of this functional is not known.

The third term, $G_{\rm ne}[v]$, is the total energy of the {\it
non-electrostatic} solute-solvent interaction given a solvent density
distribution $\rho_0v(\vec r)$. The potential $U(\vec r)=\sum_{i}
U_i(\vec r-\vec r_i)$ is the sum of the (short-ranged) repulsive
exclusion and (long-ranged) attractive dispersion interaction between
each solute atom $i$ at position $\vec r_i$ and a solvent molecule at
$\vec r$. Classical solvation studies typically represent the
interaction $U_i$ as an isotropic Lennard-Jones (LJ) potential,
\begin{eqnarray}
 U_{\rm
LJ}(r)=4\epsilon\left[\left(\frac{\sigma}{r}\right)^{12}-\left(\frac{\sigma}{r}\right)^6\right],
\label{lj}
\end{eqnarray} 
with an energy scale $\epsilon$, length scale $\sigma$, and
center-to-center distance $r$. Using the form of (\ref{lj}) implies
that $v({\vec r})$ is defined with respect to the LJ-centers of the
solvent molecules.

The last term, $G_{\rm es}[v]$, describes the total energy of the {\it
electrostatic} field and the mobile ions in the system expressed by
the local electrostatic potential $\Psi(\vec r)$ assuming linear
response of the dielectric solvent.\cite{jackson} Similar to
$\gamma(\vec r;[v])$, the position-dependent dielectric constant
$\epsilon(\vec r)=\epsilon(\vec r;[v])$ depends on the geometry of
$v(\vec r)$ with an unknown functional form. $\lambda(\vec r)$ is the
fixed charge density distribution of the solutes and the local energy
density of the {\it mobile ions} is \cite{sharp:honig,gilson}
\begin{eqnarray}
U_{\rm mi}(\vec r) = k_BT\sum_{j}\rho_j\{\exp[-\beta q_j \Psi(\vec
r)]-1\}.
\label{pb}
\end{eqnarray}
with the thermal energy $k_BT=\beta^{-1}$.  Variation of
(\ref{eq:grand}) for a fixed $v(\vec r)$ with respect to $\Psi(\vec
r)$ yields the Poisson-Boltzmann equation\cite{sharp:honig,gilson}
\begin{eqnarray}
{\rm PB}(\vec r)=0 &=& \nabla\cdot[\epsilon_0\epsilon(\vec r;[v])\nabla\Psi(\vec
r)]+\lambda(\vec r)\nonumber\\&+&v(\vec r)\sum_{j}q_j\rho_j\exp[-\beta q_j
\Psi(\vec r)],
\label{pb}
\end{eqnarray}
where $q_j$ and $\rho_j$ are the charge and concentration of the
mobile ion species $j$. Note that the ionic charge density in
(\ref{pb}) is multiplied by $v(\vec r)$ to account for the fact that
ions usually cannot penetrate the volume empty of polar solvent due to
a huge free energy penalty. We remark that the treatment of the
electrostatics in our theory has the same limitations as other
implicit models using PB, for instance when describing highly charged
or strongly correlated electrolyte systems. In contrast to PB/SA
models however, the dielectric boundary is optimized such that it
responds to the local nonpolar and polar potential; it is not assumed
beforehand.

Let $v_{\rm min}(\vec r)$ be the exclusion function which minimizes
the functional (\ref{eq:grand}).  Then, the resulting Gibbs free
energy of the system is given by $G[v_{\rm min}]$. The solvation free
energy $\Delta G$ is the reversible work to solvate the solute and is
given by
\begin{eqnarray}
\Delta G=G[v_{\rm min}]-G_0,
\end{eqnarray}
where $G_0$ is a constant reference energy which can refer to the pure
solvent state and an unsolvated solute.  The potential of mean force
(pmf) along a given reaction coordinate $x$ (e.g. the distance
between two solutes' centers of mass) is given, within a constant, by
$G[v_{\rm min}]$, where $v_{\rm min}(\vec r)$ must be evaluated for
every $x$.  In order to proceed we will need valid approximations for
$\gamma(\vec r;[v])$ and $\epsilon(\vec r;[v])$ with which $v_{\rm
min}(\vec r)$ can be calculated by explicitly minimizing our free
energy functional (\ref{eq:grand}) according to (\ref{min}).

\subsection{Approximations for $\gamma(\vec r;[v])$ and $\epsilon(\vec r;[v])$}

Let us start with a possible description of $\gamma(\vec r;[v])$.  For
a planar macroscopic interface the parameter $\gamma$ is usually
identified by the surface tension of the solvent adjacent to the
second medium. This surface tension obviously depends on the
microscopic interactions between the medium and the solvent and is
generally decreased by attractive dispersion or electrostatic
contributions.  It seems that microscopic interactions are adequately
represented by a macroscopic quantity like $\gamma$ {\it if} their
range is much smaller than the investigated length scales, such as the
radii of curvature or mean particle distances. The effect of the
microscopic interactions are then absorbed in $\gamma$. This has been
exemplified with free energy estimates for the solvation of large,
neutral plate-like or spherical alkane-assembled
solutes,\cite{huang:pnas,huang:jpcb:2002} For the description of
solvation on smaller length scales, however, it seems important to
separate the free energy into a part which accounts for the formation
of a cavity and a part which describes the dispersion interactions
explicitly.\cite{gallicchio:jpcb} Furthermore, it has been shown that
the water liquid-vapor surface tension $\gamma_{\rm lv}$ is the
asymptotic value of the solvation free energy per surface area for
hard spherical cavities in water in the limit of large
radii.\cite{lum:jpc,huang:jpc} These considerations motivate our
choice of the second and third term in the functional (\ref{eq:grand})
and lead to the assumption $\gamma\equiv\gamma_{\rm lv}$ in the limit
of vanishing curvatures.

The surfaces of realistic (bio)molecules, however, display highly
curved shapes, so $\gamma(\vec r;[v])$ will strongly depend on the interface
geometry around $\vec r$ in a complicated fashion. In the following we
make a {\it local curvature approximation}, i.e.~we assume that
$\gamma(\vec r;[v])$ can be expressed solely as a function of the local mean
curvature
\begin{eqnarray}
H(\vec r)=\left(\kappa_1(\vec r)+\kappa_2(\vec r)\right)/2=R(\vec
r)^{-1},
\end{eqnarray}
where $R(\vec r)$ is the radius of mean curvature and $\kappa_1(\vec r)$ and
$\kappa_2(\vec r)$ are the local principal curvatures of the
interface.\cite{prince} The mean curvature $H$ is only defined at the
boundary of $v(\vec r)$. We have chosen the convention in which the
curvatures are positive for convex surfaces (e.g. a spherical cavity)
and negative for concave surfaces (e.g. a spherical droplet).

The curvature dependence of the liquid-vapor surface tension
has been a long standing subject of research and is still under steady
discussion.\cite{hendersonreview,bieker,evans2} For water, which is
close to the critical point under ambient conditions, $\gamma$ is
argued to be nonanalytical in curvature.\cite{evans2}
The first order correction term, however, is likely to be linear in
curvature as predicted by scaled-particle theory,\cite{spt} the
commonly used ansatz to study the solvation of hard spherical
cavities. Although this result is only strictly valid for the case of
spherical particles, we assume that it can be applied to local mean
curvatures such that $\gamma(\vec r;[v])$ reduces to the function
\begin{eqnarray}
\gamma(\vec r)=\gamma_{{\rm lv}}(1-2\delta H(\vec r)),
\label{H}
\end{eqnarray}
where $\delta$ is the Tolman length, which is expected to be of
molecular size.\cite{tolman} In our study we assume $\delta$ is
constant and positive, while the curvature can be positive or negative
as defined above.  Note that this leads to an increase of surface
tension for concave surfaces in agreement with the geometrical
arguments of Honig {\it et al.}\cite{nicholls} in their solvation
study of alkanes. It has been shown by computer simulations of growing
a hard spherical cavity in water that (\ref{H}) predicts the
interfacial energy rather well for radii $\gtrsim
3$\AA.\cite{huang:jpc} A major drawback of approximation (\ref{H}) is
that it gives unphysical results if the radius of mean curvature is
smaller than twice the Tolman length, $|R|<2\delta$. It yields
negative and diverging surface tensions for convex and concave
surfaces, respectively.  The latter is not possible due to the finite size
of the solvent molecules. Thus, care has to be taken with approximation (\ref{H}) when investigating systems which can exhibit radii of
curvature $|R|<2\delta$.

Let us now turn to electrostatics. The most common approximation for the
position-dependent dielectric constant is proportional to the volume
exclusion function $v(\vec r)$,\cite{roux:biochem} such that the functional
$\epsilon(\vec r;[v])$ reduces to,
\begin{eqnarray}
\epsilon(\vec r)=\epsilon_v+v(\vec r)(\epsilon_l-\epsilon_v),
\label{e}
\end{eqnarray}
where $\epsilon_v$ and $\epsilon_l$ are the dielectric constants
inside and outside the volume $\cal V$, respectively.  Eq.~(\ref{e})
is valid only in the limit of large solute sizes when the molecular size
of the solvent is negligible. For charged solutes on a molecular
scale, let's say mono- or polyvalent ions, two difficulties
arise. First the electric field close to the highly curved solutes can
be strong enough for the dielectric constant to be field
dependent. This formally affects the form of the electrostatic term in
the free energy functional which assumes a linear response of the
solvent. An improvement for continuum models along these lines has
been proposed by Luo et al.\cite{luo:jpc} Second, the effective
position of the dielectric boundary is known to depend on the sign of
the solute charge for asymmetric solvent molecules like water. This
expresses itself, for instance, in different Born radii for two
equally charged ions which have exact the same LJ parameters but a
different sign of charge. A reasonable improvement of (\ref{e}) would
be to shift the dielectric boundary at $\vec r$ {\it parallel} to the
volume boundary by a potential dependent amount $\Delta \vec r=\xi(\Psi(\vec r))\vec n(\vec r)$:
\begin{eqnarray}
\epsilon(\vec r)=\epsilon_v+v\left(\vec r-\Delta \vec r\right)(\epsilon_l-\epsilon_v),
\label{e2}
\end{eqnarray}
where $\vec n$ is the unit normal vector to the interface. We do not
attempt however, to find an approximation for the function $\xi(\Psi)$
in this work and postpone this investigation to later studies. For
further illustration of our approach we content ourselves with the
approximations (\ref{H}) for $\gamma(\vec r;[v])$ and (\ref{e}) for
$\epsilon(\vec r;[v])$.

\subsection{Minimization of the free energy functional}
For the functional derivative of the interfacial term,
$G_{\rm int}[v]$, we utilize
\begin{eqnarray}
\frac{\delta}{\delta v}\int_{\cal W}{\rm d}^3r \;|\nabla v(\vec r)|=\frac{\delta}{\delta v}\int_{\partial\cal W}{\rm d}S=-2H
\end{eqnarray}
and
\begin{eqnarray}
\frac{\delta}{\delta v}\int_{\cal W}{\rm d}^3r\;H \;|\nabla v(\vec r)|=\frac{\delta}{\delta v}\int_{\partial\cal W}{\rm d}S\;H=-K,
\end{eqnarray}
which has been derived in detail by Zhong-can and Helfrich by means of
differential geometry.\cite{helfrich} The variable
$K(\vec r)=\kappa_1(\vec r)\kappa_2(\vec r)$ is the Gaussian curvature of the interface, which
is an intrinsic geometric property of $v$. Plugging in approximations
(\ref{H}) and (\ref{e}) into (\ref{eq:grand}), and minimizing with
(\ref{min}), we obtain,
\begin{eqnarray}
0&=&{\rm de}(\vec r)= P+2\gamma_{{\rm lv}}\left[H(\vec r)-\delta K(\vec
r)\right]-\rho_0 U(\vec r)\nonumber\\ &-&\frac{\epsilon_0}{2}[\nabla\Psi(\vec
r)\epsilon(\vec
r)]^2\left(\frac{1}{\epsilon_l}-\frac{1}{\epsilon_v}\right)-U_{\rm
mi}(\vec r).\nonumber\\
\label{diff}
\end{eqnarray}
Eq.~(\ref{diff}) is a partial second order differential equation (de)
for the optimal exclusion function $v_{\rm min}(\vec r)$ expressed in
terms of pressure, curvatures, short-range repulsion, dispersion, and
electrostatic terms, all of which have dimensions of energy density.
It can also be interpreted as a mechanical balance between the forces
per surface area generated by each of the particular contributions.
Thus, in our approach the surface shape and geometry, expressed by $H$
and $K$, are directly related to the inhomogeneous potential
contributions.  The constant solute charge density
$\lambda(\vec r)$ does not appear explicitly in (\ref{diff}) but is
implicitly considered in the PB equation (\ref{pb}), which must be
solved simultaneously. If curvature correction ($K$-term) and the last
three energetic terms are neglected one obtains the Laplace-Young
equation,
\begin{eqnarray}
P=-2\gamma_{lv}H,
\label{ly}
\end{eqnarray}
which is exclusively used for the shape description of macroscopic
capillary and interfacial phenomena in conjunction with appropriate
boundary conditions, e.g. prescribed liquid-solid contact angles at
the solid surfaces.\cite{kralchevsky} In our description the boundary
conditions are provided by the constraints given by the short-ranged
repulsive term in $U(\vec r)$, and the distribution of dispersion and
electrostatics, allowing an extrapolation of the Laplace-Young
description to mesoscopic and microscopic scales. Notice that in our
approach the solvent is treated as a continuum while the solute is
explicitly resolved.  One could use a coarse-grained treatment for the
solute by including the appropriate non-electrostatic and
electrostatic interactions in (\ref{eq:grand}).

The solution of (\ref{diff}) requires an appropriate
parametrization, i.e.~coordinate representation, for
the curvatures $H$ and $K$, such that the equation is expressed as a
function of the vector $\vec r$ and its first and second derivatives
in space. Analytical solutions to the much simpler Eq.~(\ref{ly}) and thus
to (\ref{diff}) are only available for systems with very simple geometries.\cite{kralchevsky}
Thus we use numerical solutions of (\ref{diff}) in the following to
further illustrate our theory.

\section{Applications}

First, we will consider the solvation of microscopic solutes such as
noble gases, simple alkanes, and ions which can be treated as neutral
or charged Lennard-Jones spheres. Then, we will investigate alkane
assemblies on a larger scale where interfacial and dewetting effects
are much more dominant.  For simplicity and a better transparency of
the results, mobile ions will be neglected in these illustrations.

\subsection{One Lennard-Jones sphere}

In this section we compare our approach to results from SPC and SPC/E
explicit solvent simulations.\cite{berendsen:jpc} We refrain from 
comparing to real experiments since approximations in computer
experiments are more easily controlled and the LJ parameters
of the solutes are commonly parametrized to yield accurate results in
classical simulations.

For a spherical solute with a charge $Q$ homogeneously distributed
over its surface, the functional (\ref{eq:grand}) with approximations
(\ref{H}) and (\ref{e}) and no mobile ions reduces to a function of
$R$, the radius of the sphere empty of solvent. The solvation free
energy is
\begin{eqnarray}
\Delta G(R)&=&\Delta G_{\rm pr}(R)+\Delta G_{\rm int}(R)+\Delta G_{\rm ne}(R)+\Delta G_{\rm es}(R)\nonumber\\
&=&\frac{4}{3}\pi R^3 P + 4\pi R^2 \gamma_{{\rm
 lv}}\left(1-\frac{2\delta}{R}\right)\nonumber\\&+& \int_R^{\infty} 4\pi r^2{\rm
 d}r\; \rho_0U_{\rm LJ}(r)\nonumber\\ &+&\frac{Q^2}{8\pi \epsilon_0
 R}\left(\frac{1}{\epsilon_l}-\frac{1}{\epsilon_v}\right)
\label{eq:sphere}. 
\end{eqnarray}
Note that the last term in (\ref{eq:sphere}) is equivalent to the Born
electrostatic solvation free energy.\cite{roux:biochem} Recently,
Manjari {\it et al.} have presented a very similar expression for the
solvation of a charged spherical cavity on the basis of a
minimization principle and have investigated the variation of $R$ with
thermodynamic conditions.\cite{kim} Differentiation of 
(\ref{eq:sphere}) with respect to $R$ and subsequent division by $4\pi
R^2$ yields
 \begin{eqnarray}
  0&=& P+\frac{2\gamma_{{\rm
  lv}}}{R}\left(1-\frac{\delta}{R}\right)-\rho_0U_{\rm
  LJ}(R)\nonumber\\&-&\frac{Q^2}{32\pi^2\epsilon_0
  R^4}\left(\frac{1}{\epsilon_l}-\frac{1}{\epsilon_v}\right)
\label{sphere2},
 \end{eqnarray}
which is in accord with Eq.~(\ref{diff}) given sphere-like curvatures,
$H=1/R$ and $K=1/R^2$.   We
can now calculate the solvation free energies of simple spherical
solutes, such as noble gases or ions.  The free parameters in
Eq.~(\ref{sphere2}) are the pressure $P$, Tolman length $\delta$,
liquid-vapor surface tension $\gamma_{\rm lv}$, and dielectric
constants $\epsilon_v$ and $\epsilon_l$.
 
\subsubsection{One neutral LJ sphere}
First, let us focus on uncharged spheres, for which the electrostatic
term in (\ref{eq:sphere}) can be neglected.  We compare the results
from our theory to those calculated by Hummer {\it et
al.}\cite{hummer:jpc:1996} for neutral LJ spheres in SPC water, and
those calculated by Paschek\cite{paschek} for noble gases in SPC and
SPC/E water.  The solute-water LJ parameters $\sigma$ and $\epsilon$
are summarized in Tab.~I.  The surface tension $\gamma_{\rm lv}$ was
set to that estimated for SPC and SPC/E water at 300K, $\gamma_{\rm
lv}=65$mJ/m$^2$ and $\gamma_{\rm lv}=72$mJ/m$^2$,
respectively.\cite{alejandre,huang:jpc} The pressure is fixed to 1atm.
Finally, the remaining free parameter $\delta$ was fit to {\it reproduce}
the simulation solvation free energies exactly.  The solvation
free energies from simulation $\Delta G_{\rm sim}$ and best fit Tolman
lengths $\delta_{\rm bf}$ are shown in Tables~I~and~II for the SPC and
SPC/E models, respectively.

Before we discuss the results, let us compare the particular energy
contributions $\Delta G_{\rm i}(R)$ with ${\rm i=pr,int,ne}$ for
Na$^0$ (plotted in Fig.~\ref{fig:G}). As anticipated, the pressure
term $\Delta G_{\rm pr}(R)$ with $P=1$atm is negligible compared to
the other contributions. The interfacial term $\Delta G_{\rm int}(R)$
increases with the cavity radius $R$. The integrated LJ-interaction
term $\Delta G_{\rm ne}(R)$ shows long-range attraction and a steep
short-ranged repulsion with a minimum at
$R=\sigma(Na^{0})=2.85$\AA. The total solvation free energy for the
Na$^0$ shows a single minimum at $R_{\rm min}=2.32$\AA~ with $\Delta
G=9.2$kJ/mol for a $\delta_{\rm bf}$=0.79\AA.

The results for the other LJ-spheres, summarized in Tab.~I and II,
reveal several noteworthy observations.  First, the best fit Tolman
lengths $\delta_{\rm bf}$ range from 0.76\AA~ to 1.00\AA~; they are
not only of molecular size, as expected, but are approximately half
the LJ-radius of a SPC or SPC/E water molecule.  Second, the
$\delta_{\rm bf}$ values for noble gases in SPC/E water (Tab.~II) are
approximately 10$\%$ larger than those in SPC water (Tab.~I).  This is
in qualitative agreement with the results of Huang {\it et al.} who
measured $\delta$=$0.76\pm0.05$\AA~ and $\delta$=$0.90\pm0.03$\AA~ for
SPC and SPC/E, respectively, by fitting Eq.~(\ref{H}) to the hydration
free energy of hard spheres with varying radii.\cite{huang:jpc}

Third, the quite accurate data of Paschek demonstrate a systematic increase of 
$\delta_{\rm bf}$ with solute size.  The inability of our theory 
to be fit by one fixed constant $\delta_{\rm bf}$ points to the 
anticipated fact that Eq.~(\ref{H}) can not capture strong curvature 
effects accurately and will have to be refined for small solutes. Despite this shortcoming,
these results show surprisingly good agreement; if we assume a fixed
delta, for instance $\delta=0.91$\AA~ for all noble gases in the SPC
data of Paschek, our theory predicts results within
$15\%$ of the simulation data. Finally, we
observe that the effective optimal sphere radius $R_{\rm min}$ is
always smaller than the radius of the canonical SAS with a
typical probe radius of 1.4\AA,\cite{connolly} $R_{\rm
min}<(\sigma_{ss}/2+1.4{\rm \AA})\simeq\sigma$, but larger than the
vdW surface, $R_{\rm min}>\sigma_{ss}/2$, where
$\sigma_{ss}$ is the solute-solute LJ-length.\cite{sigmanote}

\subsubsection{One charged LJ sphere}
Let us now turn to charged Lennard-Jones spheres (ions) also examined
in the paper by Hummer {\it et al.}  with SPC water simulations. We
assume $\delta$ to be the {\it fixed} by the previously obtained
$\delta_{\rm bf}$ values for uncharged spheres.  The dielectric
constants are set to $\epsilon_v=1$ and $\epsilon_l=65$, in accord
with SPC water.\cite{spoel} The electrostatic contribution $\Delta
G_{\rm es}(R)$ and the total $\Delta G(R)$ for Na$^+$ are shown in the
inset of Fig.~\ref{fig:G}. The electrostatic contribution decreases
the optimal radius to $R_{\rm min}=1.83$\AA~ giving a solvation free
energy of $\Delta G=-334$kJ/mol. In fact, the optimal sphere radius
$R_{\rm min}$ is always considerably smaller for the charged solutes
(Tab.~III) than for their neutral counterparts (Tab.~I). This is
caused by the strong compressing force of the polar solvent attempting
to penetrate the low dielectric cavity.  The solvation free energies
from theory $\Delta G$ and those from simulation $\Delta G_{\rm sim}$
are also shown in Tab.~III.  While our theory describes the hydration
free energies for positively charged ions within $15\%$, it
considerably underestimates those of the negative ions. This
qualitative disagreement between positive and negative ions was
expected since the Born radii for anions are always smaller than those
for cations, a consequence of the different solvation structure around
charged solutes with opposite signs. As mentioned in the previous
section, the position of the dielectric boundary has to be refined for
accurate estimates of the electrostatic contribution to the hydration
free energy. If we apply the correction (\ref{e2}) to the dielectric
boundary with a simple, potential-independent shift $\xi_+=-0.25$\AA~
for positive and $\xi_-=-1.05$\AA~ for negative spheres such that the
dielectric boundary has a radius of $R+\xi_\pm<R$, improved values
($\Delta G_\xi$ in Tab.~III) are obtained which reproduce all
simulation values within $10\%$!

\subsection{Linear alkanes}

Let us now consider simple polyatomic molecules, such as ethane,
propane, or butane in a cylindrically symmetric one-dimensional (1D)
chain conformation.  Other conformations will be neglected. The
symmetry of these systems allows us to express the volume exclusion
function $v(\vec r)$ of the enveloping surface by a one dimensional
shape function $r(z)$, where $z$ is the coordinate on the symmetry
axis and $r$ the radial distance to it. The full surface in
three-dimensional space is obtained by revolving the shape function
$r(z)$ around the symmetry axis. Technical details are given in the
Appendix.

The LJ parameters for ethane and methane are the same as those used by
Ashbaugh {\it et al.}\cite{ashbaugh:biophys} in their SPC simulation
of linear alkanes (see Tab.~I).  The simulation solvation energy of
the spherical methane, $\Delta G=10.96$kJ/mol, can be reproduced with
$\delta_{\rm bf}=0.85$\AA. Solving the cylindrically symmetric problem
for ethane using the same $\delta$, we obtain a fit-parameter free
$\Delta G=11.40$kJ/mol, which is only 7$\%$ larger than the simulation
results. Alternatively, $\delta_{\rm bf}=0.87$\AA~ reproduces the
simulation energy exactly.  This is surprisingly good agreement
considering the crudeness of our curvature correction and the fact
that the large curvature of the system varies locally in space. The
curvature and shape functions are plotted in Fig.~\ref{fig:ethane}
together with the vdW surface and the canonical SAS obtained from
rolling a probe sphere with the typically chosen radius $r_{\rm
p}=1.4$\AA~ over the vdW surface.\cite{connolly} Away from the
center of mass $|z|\gtrsim 1$\AA~ the curvatures follow the expected
trends for the spherical surfaces: $H= 1/R$ and
$K=1/R^2$ with $R \simeq 3.1$\AA~.  The
optimal surface resulting from our theory is smaller than the
canonical SAS and smooth at the center of mass ($z=0$) where the
canonical SAS has a kink. Thus our surface has a smaller mean
curvature at $z=0$ and an almost zero Gaussian curvature, which is
typical for a cylinder geometry with one principal curvature equal to
zero.  These results may justify the use of smooth surfaces in
coarse-grained models of closely-packed hydrocarbons, a possibility we
will explore in the following section with solvation on larger length
scales where dewetting effects can occur.  If we repeat the above
calculation for propane and butane (three and four LJ-spheres, see
also Tab.~I for parameters) we need $\delta_{\rm bf}=0.94$\AA~ and
$\delta_{\rm bf}=0.96$\AA, respectively, to reproduce the simulation
results exactly. The increasing difference in $\delta_{\rm bf}$
compared to methane and ethane is likely due to contributions from
other than cylindrically symmetric conformations which were ignored in
our analysis.

\subsection{Two spherical nanosolutes}

\subsubsection{Model}

Let us now consider two spherical solutes which represent
homogeneously assembled CH$_2$ groups with a uniform density
$\rho$=0.024\AA$^{-3}$ up to a radius $R_0=15$\AA, defined by the
maximal distance between a CH$_2$ center and the center of the solute.
Integration of the CH$_2$-water LJ interaction over the entire sphere
yields a 9-3 like potential $U_i(r)$ for the interaction between the
center of the solute ($i=1,2$) and a water
molecule. \cite{huang:jpcb:2002} The intrinsic, nonelectrostatic
solute-solute interaction $U_{\rm ss}$ can be obtained in a similar
fashion.  The CH$_2$-water LJ parameters, $\epsilon=0.5665$kJ/mol and
$\sigma=3.52$\AA, are taken from the OPLSUA
force-field\cite{jorgensen} and are similar to those used by Huang
{\it et al.} in their study on dewetting between paraffin
plates.\cite{huang:jpcb} Minimizing Eq.~(\ref{eq:sphere}) for just one
sphere we obtain an optimal solvent excluded radius of $R_{\rm min}
\simeq 17.4$\AA, which is $R_0+2.4$\AA. Since we are also
interested in the effects of electrostatic interactions we place
opposite charges $\pm Ze$, where $e$ is the elementary charge, in the
center or on the edge of the two spheres. Poisson's equation is
simultaneously solved on a two-dimensional grid in cylindrical
coordinates. Numerical details are given in the Appendix.

The solvation of the two solutes is studied for a fixed
surface-to-surface distance which we define as $s_0=r_{12}-2R_0$,
where $r_{12}$ is the solute center-to-center distance. The effective
surface-to-surface distance defined by the accessibility of the
solvent centers is thus $s\simeq r_{\rm 12}-2R_{\rm
min}=s_0-4.8$\AA. In the following we focus on a separation distance
of $s_0=8$\AA~ to investigate the influence of different energetic
contributions on the shape function, $r(z)$, and the curvatures,
$K(z)$ and $H(z)$. For $s_0 = 8$\AA, it follows that $s\simeq 3.2
$\AA, such that two water molecules could fit between the solutes on
the $z$-axis. We systematically change the solute-solute and
solute-solvent interactions, as summarized in Tab.~I. We begin with
only the repulsive part of the nonelectrostatic interaction $U_i(r)$
in system I, and then add a curvature correction with
$\delta=0.75$\AA, vdW attractions, and sphere-centered charges $Z = 4$
and $Z = 5$ in systems II-V, respectively.  To study the influence of
charge location, we reduce the magnitude of each charge in system VI
to $Z=1$ and move them to the edge of the spheres on the symmetry axis
such that they are $8$\AA~apart (indicated by arrows in
Fig.~\ref{fig:nano}.  The surface tension and dielectric constants of
the vapor (solute) and liquid are fixed to $\gamma_{\rm
lv}=72$mJ/m$^2$, $\epsilon_v=1$, and $\epsilon_l=78$, respectively.

\subsubsection{Behavior of the shape function}

The results for the curvatures and shape function, defined by $r(z)$,
for systems I-VI are shown in Fig.~\ref{fig:nano}. Away from the
center of mass ($|z|\gtrsim 10$\AA), systems I-VI show very little
difference. The curvatures are $H=1/R$ and $K=1/R^2$ with $R \simeq
17.4$\AA. Close to the center of mass ($z\simeq 0$), however, the
influence of changing the parameters is considerable. In system I,
Eq.~(\ref{diff}) reduces to the minimum surface equation $H(z)=0$ for
$z\simeq 0$. For two adjacent spheres the solution of this equation is
the catenoid $r(z)\simeq{\rm cosh}(z)$, which features zero mean
curvature ($\kappa_1$ and $\kappa_2$ cancel each other) and negative
Gaussian curvature. As a consequence, the system exhibits a vapor
bubble bridging the solutes, i.e. water is removed from the region
between the spheres even though it fits there. This dewetting is
driven by the interfacial term $G_{\rm int}$ which always favors
minimizing the liquid-vapor interface.

When curvature correction is applied (system II), the mean curvature
becomes nonzero and negative (concave) at $z\simeq 0$, while the
Gaussian curvature grows slightly more negative. Thus the total
enveloping surface area becomes larger and the solvent inaccessible
volume shrinks, i.e. the value of the shape function at $z \simeq 0$
decreases. Turning on solute-solvent dispersion attraction amplifies
this trend significantly as demonstrated by system III. Mean and
Gaussian curvatures increase fivefold, showing strongly enhanced
concavity, and the volume empty of water decreases considerably,
expressed by $r(z=0)\simeq 10.7$\AA~ dropping to $r(z=0)\simeq
6.3$\AA. These trends continue with the addition of electrostatics in
system IV.  When the sphere charges are further increased from $Z=4$
to $Z=5$ (system IV$\rightarrow$V), we observe a wetting transition:
the bubble ruptures and the shape function jumps to the solution for
two isolated solutes, where $r(z\simeq 0)=0$. The same holds when
going from III to VI, when only one charge unit, $Z=1$, is placed at
each of the solutes' surfaces. Importantly, this demonstrates that the
present formalism captures the sensitivity of dewetting phenomena to
specific solvent-solute interactions as reported in previous
studies.\cite{huang:jpcb,pettitt,dzubiella:channel1,vaitheesvaran,dzubiella:jcp:2003,zhou:science,berne:nature}
Note that the optimal shape function at $|z|\simeq\pm2$\AA~ is closer
to the solutes in VI compared to V due to the proximity of the charge
to the interface. Clearly, the observed effects, particularly the
transition from III to VI, cannot be described by existing solvation
models which use the surface area (GB/SA or PB/SA)\cite{roux:biochem}
or effective surface tensions and macroscopic solvent-solute contact
angles\cite{kralchevsky,attard} as input.

\subsubsection{Potential of mean force}
The significant change of the shape function with the
solute-solvent interaction has a strong impact on the potential of
mean force (pmf) (or effective interaction) between the solutes
\begin{eqnarray}
W(s_0)= G(s_0)- G(\infty)+U_{\rm ss}(s_0).
\end{eqnarray} 
Recall that $U_{\rm ss}$ is the instrinsic dispersion interaction
between the two solutes. Values of $W(s_0=8{\rm\AA})$ are given in
Tab.~IV. From system I to VI the total attraction between the solutes
decreases almost two orders of magnitude. Interestingly, the curvature
correction (I$\rightarrow$II) lowers $W$ by a large 23.5$k_BT$, even
though $R_0\gg\delta$. The reason is that the mean radii of curvature
between the spheres can assume values $\simeq \delta$, implying that
curvature correction is also important for large solutes. A striking
effect occurs when vdW contributions are introduced
(II$\rightarrow$III): the inter solute attraction decreases by
approximately $28k_BT$ while the dispersion solute-solute potential
$U_{\rm ss}(s_0=8{\rm \AA})$ changes by only $-0.44 k_BT$. Similarly,
adding charges of $Z=5$ (III $\rightarrow$ V) at the solutes' centers
or $Z=1$ (III $\rightarrow$ VI) at the solutes' surfaces decreases the
total attraction by 1.2$k_BT$ and 5k$_B$T, respectively. Note that the
total attraction {\it decreases} even though electrostatic attraction
has been added between the solutes. The same trend has been observed
recently in explicit water simulations of a similar system of charged
hydrophobic nanosolutes.\cite{dzubiella:jcp:2003,dzubiella:jcp:2004}

Now we turn our attention to varying the intersolute distance.  The
pmfs and solute-solute mean forces $F=\partial W(s_0)/\partial s_0$
versus a range of $s_0$ are shown for systems I,II,III, and VI in 
Fig.~\ref{fig:pmf}. System I, with purely repulsive
solute-solvent interactions, displays a strong attraction
($W\simeq-150k_BT$) at $s_0=2$\AA~ which decreases, almost linearly,
to zero at a distance $s_0=13.5$\AA~ where the system shows a wetting
transition. The corresponding force is discontinuous at this critical
distance.  The steep repulsion at short intersolute distances
($s_0\simeq 1.5$\AA~) stems from the repulsive term of the LJ
interaction between the solutes. Note that the intrinsic solute-solute
interaction $U_{\rm ss}(s_0)$, also shown in Fig.~\ref{fig:pmf}, is almost two
orders of magnitude smaller than the hydrophobic attraction. Adding
the curvature correction in system II decreases the range and strength
of the pmf by approximately $20\%$, which is significant and
unexpected since $R_0\gg\delta$.  Adding dispersion
attractions in system III decreases the range and strength of the
hydrophobic attraction considerably, but it is still much stronger
than the inter solute dispersion attraction $U_{ss}$ alone. When
surface charges ($Z=1$) are added in system VI, the range of
hydrophobic attraction further decreases but the total attraction
increases at short intersolute distances.  This is due to the
increasing size of the bridging bubble ($r(z=0)$ increases) as the two
solutes approach each other, which decreases the high dielectric
screening of the solute-solute electrostatic attraction.  This again
underlines the importance of coupling electrostatics and dewetting
effects, as the electrostatic attraction (or repulsion) may be
magnified by more than an order of magnitude when dewetting
occurs. For charges with opposite sign this could be interpreted as the
stabilization of a salt bridge due to dehydration.\cite{fernandez}
Systems IV and V, not shown in Fig.~\ref{fig:pmf}, exhibit the same
qualitative behavior as system VI.

\subsubsection{Comparison of mean forces to previous MD simulations}
We continue considering the mean force between two nanosized solutes
and compare our theory now to the MD simulations of Dzubiella {\it et
al.}\cite{dzubiella:jcp:2003,dzubiella:jcp:2004} Their solute model
slightly differs from the one used in the previous section: the
solute-solvent interaction potential is purely repulsive and is given
by $U_i(r)=k_BT(r-R_0)^{-12}$, while the solute-solute interaction is
hard-sphere like with a hard sphere radius $R_0$. The solutes are
neutral or carry opposite charges $Q$ homogeneously distributed over
the sphere volume. The simulations were carried out with the SPC/E
model of water. In our theory, we fix the Tolman length to
$\delta=0.90$\AA~ as measured by Huang {\it et al.}\cite{huang:jpc}
and the dielectric constant to $\epsilon_l=71$ for SPC/E
water.\cite{berendsen:jpc} The mean forces are shown in
Fig.~\ref{fig:mf} for neutral spheres of radii $R_0=10$ and 12\AA~ and
oppositely charged solutes with radius $R_0=10$\AA~ and charge $Q=2$e
and 5e versus the solutes' surface-to-surface distance $s_0$.
Simulation and theory are in good, almost quantitative agreement, and
show that our theory captures the decreasing range of the strongly
hydrophobic attraction with decreasing radius and increasing
charge due to suppressed dewetting. We emphasize that our theory is
basically fit-parameter free for this system of large
solutes. Fig.~\ref{fig:mf} also shows the theoretical mean force for
the neutral $R_0=12$\AA~ solute using a smaller Tolman length
$\delta=0.75$\AA. The decrease of the Tolman length increases the
depth and range of the solvent-mediated solute-solute attractive mean
force by approximately 5\%, showing a nonvanishing but only slight
influence.

\section{Conclusion and final remarks}

In summary, we have presented a novel implicit solvent model which
couples polar and nonpolar solvation contributions by employing a
variational formalism in which the Gibbs free energy of the system is
expressed as a functional of the solvent volume exclusion
function. Minimization of the free energy leads to a Laplace-Young
like equation for the solvent excluded cavity around the solutes,
which is extended to describe solvation on mesoscopic and microscopic
scales.  We have shown that the theory gives a reasonable description
of the solvation of microscopic solutes, such as ions and
alkanes. Improved accuracy will require further refinement of the
curvature dependence of the surface tension $\gamma(\vec r;[v])$ and
the definition of the position-dependent dielectric constant
$\epsilon(\vec r;[v])$. Given the physically reasonable values of the
parameters $\delta$ and $\xi$ we found by fitting, we hope that
extensions based on physical rational, e.g. given by complementary
microscopic
approaches\cite{beglov:jcp,zwanzig,lum:jpc,hummer:pnas,garde:prl} and
further empirical corrections, will lead to an accurate fit-parameter
free implicit solvent description.

We have further demonstrated that on larger scales, where solvent
dewetting can play an important role in solvation, our formalism
captures the delicate balance between hydrophobic, dispersive and
electrostatic forces which has been observed in previous
systems.\cite{huang:jpcb,pettitt,dzubiella:channel1,vaitheesvaran,dzubiella:jcp:2003,zhou:science,berne:nature}
The dewetting in our model is driven by the interfacial term which
favors minimizing the solute-solvent interface. A comment must be made
here regarding the sensitivity of dewetting to the particular
solvent-solute interactions. As recently argued by
Chandler,\cite{chandler:review} extended fluid interfaces near phase
coexistence are often referred to as 'soft' because they can be
deformed with only little or no free-energy change.\cite{safran} Our
approach seems to account for this sensitivity since small changes of
the constraints in the differential equation (\ref{diff}) for the
shape function, given e.g. by the dispersion potential close to the
solute surface, can lead to a major deformation or even rupture
(wetting transition) of the inter-solute, dewetted region. As we have
shown, this can significantly change the pmf for the solutes. Thus we
anticipate that slight changes in the geometry of a system, e.g. a
slight concave or convex bending of two plate-like
solutes,\cite{huang:jpcb,pettitt} can lead to very different results
for the dewetting magnitude and the pmf.

The current illustrations utilized spherical and cylindrical
symmetries.  More complex molecules, such as proteins, will require
solving the full three dimensional problem.  Numerical algorithms for
the calculation of interface evolution for more complicated geometries
are provided by efficient level-set methods or fast marching
methods.\cite{level} We believe that in the full three-dimensional
(3D) case, our method will be much more efficient than other
microscopic approaches which partly resolve the water structure and
are able to describe dewetting effects, e.g. the Lum-Chandler-Weeks
theory\cite{lum:jpc} (LCW) or information theory
\cite{hummer:pnas,garde:prl} (IT), as only a two-dimensional surface
is sought rather than a 3D density distribution on a fine grid. We
remark that LCW and IT do not consider electrostatic interactions and
may benefit from our complementary approach.

\section*{Acknowledgment}

The authors thank Tushar Jain, John Mongan, and Cameron Mura for
useful discussions.  J.D. and J.M.J.S acknowledge financial support
from a DFG Forschungsstipendium and the PFC-sponsored Center for
Theoretical Biological Physics (Grants No. PHY-0216576 and
PHY-0225630), respectively. Work in the McCammon group is additionally
supported by NIH, HHMI, NBCR, and Accelrys, Inc.

\section*{Appendix A: Curvatures in cylindrical coordinates}
In our general parametrization for the shape function $r(z)$ we
express the radial coordinate by $r=r(l)$ and the axial coordinate by
$z=z(l)$, as functions of the parameter $l$. Depending on the geometry
of the considered system, $l$ has to be conveniently chosen, for
instance to be the arc length, or $r(l)=l$, or $z(l)=l$. In our
illustration the most convenient choice is $z(l)=l$. The principal
curvatures are generally given by\cite{frankel}
\begin{eqnarray}
\kappa_1(r,z)=\frac{-z'}{r\sqrt{z'^2+r'^2}} \;{\rm ,}\;\kappa_2(r,z)=\frac{z'r''-z''r'}{(z'^2+r'^2)^{3/2}},
\end{eqnarray}
where the primes indicate the partial derivative with respect to
$l$. Additionally, the unit normal vector reads
\begin{eqnarray}
\vec
n(r,z)=\frac{1}{\sqrt{z'^2+r'^2}}\left(\begin{array}{c}z'\\-r'\end{array}\right).
\end{eqnarray}
The differential equation (\ref{diff}) is then solved by a forward
relaxation scheme in time $t$
\begin{eqnarray}
\left(\begin{array}{c}r(t+\Delta t)\\z(t+\Delta
t)\end{array}\right)=\left(\begin{array}{c}r(t)\\z(t)\end{array}\right)-\Delta
t\, {\vec n}(r,z){\rm de}(r,z),
\end{eqnarray}
where the steady-state solution $\partial(r,z)/\partial t=0$ is the
solution of ${\rm de}(r,z)=0$ we are looking for.  In the numerical
calculations we use a grid of 500 bins and an integration time step of
$\Delta t=$0.001.  The first and second derivatives are approximated
using a symmetric two and three-step finite difference equation,
respectively. Convergence is usually reached after 10$^5$
time steps. The result is observed to be independent of the initial
choice of $r(z)$ at $t=0$.

\section*{Appendix B: Numerical solution of the PB equation}
Since we neglect mobile ions in our work, the PB equation reduces to
Poisson's equation. It is solved on a two dimensional grid in
cylindrical coordinates $r$ and $z$ with a finite difference method.
The gradient and Laplacian are given then by $\nabla=(\partial r,
\partial z)$ and $\Delta=\partial_r+\partial_r/r+\partial_z^2$,
respectively. The first and second derivatives are approximated using
symmetric two or three-step finite-difference equations. An explicit,
forward time relaxation scheme is used to find the solution of
Poisson's equation:
\begin{eqnarray}
\Psi(t+\Delta t;\vec r)=\Psi(t;\vec r)-\Delta t {\rm PB}(\Psi(t;\vec r)).
\end{eqnarray}
 In most cases we use a lattice spacing of $\Delta r=\Delta z=0.4$\AA~
on a $n_r\times n_z=100\times 200$ grid, and an integration time step
$\Delta t=0.05$. Convergence takes approximately $10^5$ time
steps. For the charged particles which are buried in the nanosolutes
we use homogeneously charged spheres with a radius of 2\AA.  Instead
of a sharp transition for the dielectric boundary ~(\ref{e}), we use a
smoothing function for reasons of numerical stability:
\begin{eqnarray}
\epsilon(\vec r)=\frac{\epsilon_l-\epsilon_v}{\exp(\kappa d(\vec r))+1}+\epsilon_v,
\label{e3}
\end{eqnarray}
where the absolute value of the length $d(\vec r)$ is given by the nearest
distance to the boundary of the volume exclusion function $v(\vec
r)$. $d$ is defined to be positive when $\vec r\in \cal V$
and negative elsewhere. The inverse length $\kappa$ defines the width
of the boundary region and in the limit $\kappa\rightarrow \infty$ we
recover the sharp transition (\ref{e}).  We choose a value $\kappa
\gtrsim 3{\rm \AA}^{-1}$ for which the solution of Poisson's equation
becomes basically independent of the choice of $\kappa$. An example for the
dielectric boundary is shown in Fig.~\ref{eps} for two partly dewetted
nanosolutes of radius $R_0=15$\AA~ at a distance $s_0=7$\AA~ carrying
a charge $Q=5e$ (system V in sec. III.C). 

In order to obtain the optimal shape function $v_{\rm min}(\vec r)$
the shape equation (\ref{diff}) has to be solved simultaneously with
Poisson's equation when the solutes are charged. In practice, we
first solve (\ref{diff}) without any electrostatic contributions. In
the second step, we solve  Poisson's equation with the dielectric
boundary (\ref{e3}) given by the volume exclusion function of the
former solution. The result for the electric energy density is then
plugged back in the shape equation in the third step. The last two
steps are repeated until the solution for $v_{\rm min}(\vec r)$ is
fully converged. Since the results for $r(z)$ excluding and including
electrostatics are quite similar for our systems, full convergence
takes usually only 6 to 7 repetitions of the described iteration
steps.

\newpage

\begin{table}
\begin{center}
\begin{tabular}{l | c c c | c c | c }
Solute & $\epsilon/({\rm kJ/mol})$\; & $\sigma/$\AA\; &
$\Delta G_{\rm sim}/({\rm kJ/mol})$ & $\delta_{\rm bf}/$\AA &
$R_{\rm min}/$\AA \\  \hline \hline SPC & 0.65 & 3.17 & -- & -- & --
\\ SPC/E & 0.65 & 3.17 & -- & -- & -- \\ \hline \hline
ref\cite{hummer:jpc:1996} & & & & & \\  Na$^0$ & 0.2005 & 2.85 &
9.2(1) & 0.79 & 2.32 \\ K$^0$ & 0.0061 & 4.52 & 23.7(5) & 0.76 & 2.83
\\ Ca$^{0}$ & 0.6380 & 3.17 & 10.2(3) & 0.85 & 2.80 \\ F$^0$ & 0.5538
& 3.05 & 9.7(2) & 0.85 & 2.68 \\ Cl$^0$ & 0.5380 & 3.75 & 21(3) & 0.80
& 3.30 \\  Br$^0$ & 0.4945 & 3.83 & 24(3) & 0.77 & 3.35 \\  \hline
\hline  ref\cite{paschek} & & & & & \\  Ne & 0.3156 & 3.10 &
11.41(0.05)& 0.84 & 2.61 \\  Ar & 0.8176 & 3.29 & 8.68 (0.08) & 0.90 &
2.96\\  Kr & 0.9518 & 3.42 & 8.12 (0.1) & 0.91 & 3.10 \\  Xe & 1.0710
& 3.57 & 7.65 (0.15) & 0.92 & 3.27 \\ \hline  \hline
ref\cite{ashbaugh:biophys} & & & & & \\ 
Me & 0.8941 &3.44 & 10.96 (0.46) & 0.85 & 3.11 \\  
\hline 
ethane & -- & -- & 10.75 (0.50) & 0.87 & -- \\
CH$_3$ & 0.7503 & 3.46&-- & -- & -- \\  
\hline 
propane & -- & -- & 13.81 (0.54) & 0.94 &-- \\ 
butane & -- & -- & 14.69 (0.54) & 0.96 &-- \\ 
CH$_2$ & 0.5665& 3.52 & -- & -- & -- \\ 
CH$_3$ & 0.6900 & 3.52 & -- & -- & -- \\
\end{tabular}
\caption{Solute-water LJ parameters and solvation free energy $\Delta
 G_{\rm sim}$ for neutral Lennard-Jones spheres from the SPC water
 simulations performed by Hummer {\it et al.}\cite{hummer:jpc:1996}
 and Paschek.\cite{paschek} $\delta_{\rm bf}$ is the Tolman length
 best fit to $\Delta G_{\rm sim}$ (rounded to two digits after the
 decimal point). $R_{\rm min}$ is the resulting optimal radius
 excluded of solvent. Also shown are the values for the simple alkanes
 methane (Me), ethane, propane, and butane from the study of Ashbaugh
 {\it et al.}\cite{ashbaugh:biophys} Simulation errors are given in
 parentheses.}
\label{tab1}
\end{center}
\end{table}

\begin{table}
\begin{center}
\begin{tabular}{l | c c c}
   Atom & $\Delta G_{\rm sim}/({\rm kJ\,mol}^{-1})$ & $\delta_{\rm
bf}/$\AA & $R_{\rm min}/$\AA \\  \hline Ne & 11.65 (0.05) &0.88 & 2.60
\\ Ar & 8.83 (0.08) &0.96 & 2.94 \\ Kr & 8.20 (0.1) &0.98 & 3.09 \\ Xe
& 7.58 (0.15) &1.00 & 3.25 \\
\end{tabular}
\caption{Solvation free energies for neutral Lennard-Jones spheres in
SPC/E water from the simulations of Paschek.\cite{paschek}
$\delta_{\rm bf}$ and $R_{\rm min}$ are defined as in Tab.~I.}
\label{tab1}
\end{center}
\end{table} 

\begin{table}
\begin{center}
\begin{tabular}{l | c c c c c}
  Ion & $q$ & $\frac{\Delta G_{\rm sim}}{{\rm kJ\,mol}^{-1}}$ & $\frac{\Delta G}{{\rm kJ\,mol}^{-1}}$ & $\frac{\Delta G_{\xi}}{{\rm kJ\,mol}^{-1}}$ & $R_{\rm min}/$\AA \\ 
\hline 
Na$^+$ & 1 &-398 &-334     &-394 &1.83 \\ 
K$^+$ & 1 & -271 &-246     &-282 &2.35 \\ 
Ca$^{2+}$ & 2 &-1306 &-1181&-1364 &2.09 \\ 
F$^-$ & -1 & -580 &-274    &-630 &2.25 \\ 
Cl$^-$ & -1 &-371 &-198    &-342 &2.97 \\ 
Br$^-$ & -1 & -358 &-192   &-328 &3.02 \\
\end{tabular}
\caption{Solvation free energies for charged LJ spheres in SPC water
from the simulations of Hummer {\it et al.}\cite{hummer:jpc:1996}
compared to the theoretical result $\Delta G$. For $\delta$ we use the
best fits $\delta_{\rm bf}$ to the solvation of neutral spheres as
shown in Tab.~I. $\Delta G_{\xi}$ is the result when the dielectric
boundary shift $\xi$ is applied, see text.}
\label{tab1}
\end{center}
\end{table}

\begin{table}
\begin{center}
\begin{tabular}{l | c c c c c}
  System & $\delta/{\rm \AA}\;\;\;$ & vdW attraction & $\;\;\;Z$ & $W(s_0)/k_BT$ & dewetted\\ 
\hline 
I   & 0.00 & no & 0 & -57.6 &yes \\ 
II  & 0.75 & no & 0 & -34.1 &yes \\ 
III & 0.75 & yes & 0 & -6.3 &yes \\ 
IV  & 0.75 & yes & 4 & -9.2 &yes \\ 
V   & 0.75 & yes & 5 & -5.1 &no \\ 
VI  & 0.75 & yes & 1 (oc)& -1.3 &no \\
\end{tabular}
\caption{Studied systems for two alkane-assembled spherical
solutes. $W(s_0)$ is the inter-solute pmf. If $r(z=0)\neq0$ the system
is 'dewetted'. In system VI the solutes' charge is located off-center
(oc) at the solute surface.}
\label{tab4}
\end{center}
\end{table}

\clearpage 

 \begin{figure}[htb]
 \begin{center}
   \caption{The particular solvation energy contributions $\Delta
   G_i(R)$ with $i={\rm p,int,ne}$ in Eq.~(\ref{eq:sphere}) for one LJ
   sphere with Na$^0$ parameters given in Tab.~I. The pressure term
   $\Delta G_{\rm pr}$ (thin solid line) with $P=1$atm is basically
   zero on this scale. The interfacial term $\Delta G_{\rm int}(R)$
   (dotted line) with $\gamma_{\rm lv}=65$mJ/m$^2$ increases with
   radius $R$. The LJ term $\Delta G_{\rm ne}(R)$ is given by the
   dashed line. The sum of the three contribution gives the total
   $\Delta G(R)$ (solid line) with a minimum at $R_{\rm min}=2.32$\AA~
   for the uncharged sodium Na$^0$. The inset shows the electrostatic
   contribution $\Delta G_{\rm es}(R)$ (dot-dashed line) and the total
   $\Delta G(R)$ for the charged Na$^+$ with a minimum at $R_{\rm
   min}=1.83$\AA. The best-fit Tolman length is $\delta_{\rm
   bf}=0.79$\AA.}
 \label{fig:G}
 \end{center}
 \end{figure}

 \begin{figure}[htb]
 \begin{center}
   \caption{Mean $H(z)$ and Gaussian $K(z)$ curvature and shape
   function $r(z)$ (solid lines) for ethane. The canonical SAS (dashed
   line) from rolling a probe sphere with radius $r_{\rm p}=1.4$\AA~ over
   the vdW surface (shaded region) is also shown. The vdW surface is
   defined by the solute-solute LJ-radius $\sigma_{ss}/2=1.73$\AA.\cite{sigmanote}}
 \label{fig:ethane}
 \end{center}
 \end{figure}

 \begin{figure}[htb]
 \begin{center}
   \caption{Mean $H(z)$ and Gaussian $K(z)$ curvatures and shape
   function $r(z)$ for two alkane-assembled solutes of radius
   $R_0=15$\AA~ (shaded region) at a distance $s_0=8$\AA~ for systems
   I-VI. The position of the charges $Z=\pm 1$ in VI are indicated by
   arrows. Curvatures are not shown for the 'wet' systems V and VI. }
 \label{fig:nano}
 \end{center}
 \end{figure}

 \begin{figure}[htb]
 \begin{center}
   \caption{Top frame: theoretical pmfs for the systems I-III, and VI
   versus the solute distance $s_0$. Bottom frame: corresponding mean
   forces.}
 \label{fig:pmf}
 \end{center}
 \end{figure}

\begin{figure}[htb]
 \begin{center}
   \caption{Mean force $\beta F$\AA~ between to nanosized solutes
   versus surface-to-surface distance $s_0$. The symbols denote the MD
   simulation results from Dzubiella {\it et
   al.}\cite{dzubiella:jcp:2003,dzubiella:jcp:2004} for neutral
   spheres with radius $R_0=12$\AA~ (circles) and $R_0=10$\AA
   (squares), and oppositely charged spheres with radius $R_0=10$\AA~
   and charge $Q=2$e (diamonds) and $Q=5$e (triangles). The
   corresponding theoretical results using $\delta=0.9$\AA~ are shown
   by solid lines; the range of the strong hydrophobic attraction
   decreases with decreasing radius and increasing charge. Dotted
   lines through the symbols are guides for the eye. The dashed line
   is the theory for $R_0=12$\AA~ and $\delta=0.75$\AA.}
 \label{fig:mf}
 \end{center}
 \end{figure}

 \begin{figure}[htb]
 \begin{center}
   \caption{Distribution of the dielectric constant in space for two
   nanosolutes with $R_0=15$\AA~ at a distance $s_0=7$\AA~ carrying a
   charge $Q=5e$ (system V). The region between the spheres is
   dewetted. The distribution is scaled by $\epsilon_l=78$.}
 \label{eps}
 \end{center}
 \end{figure}

\clearpage

{\Large Figure 1}
 \begin{figure}[htb]
 \begin{center}
    \epsfig{file=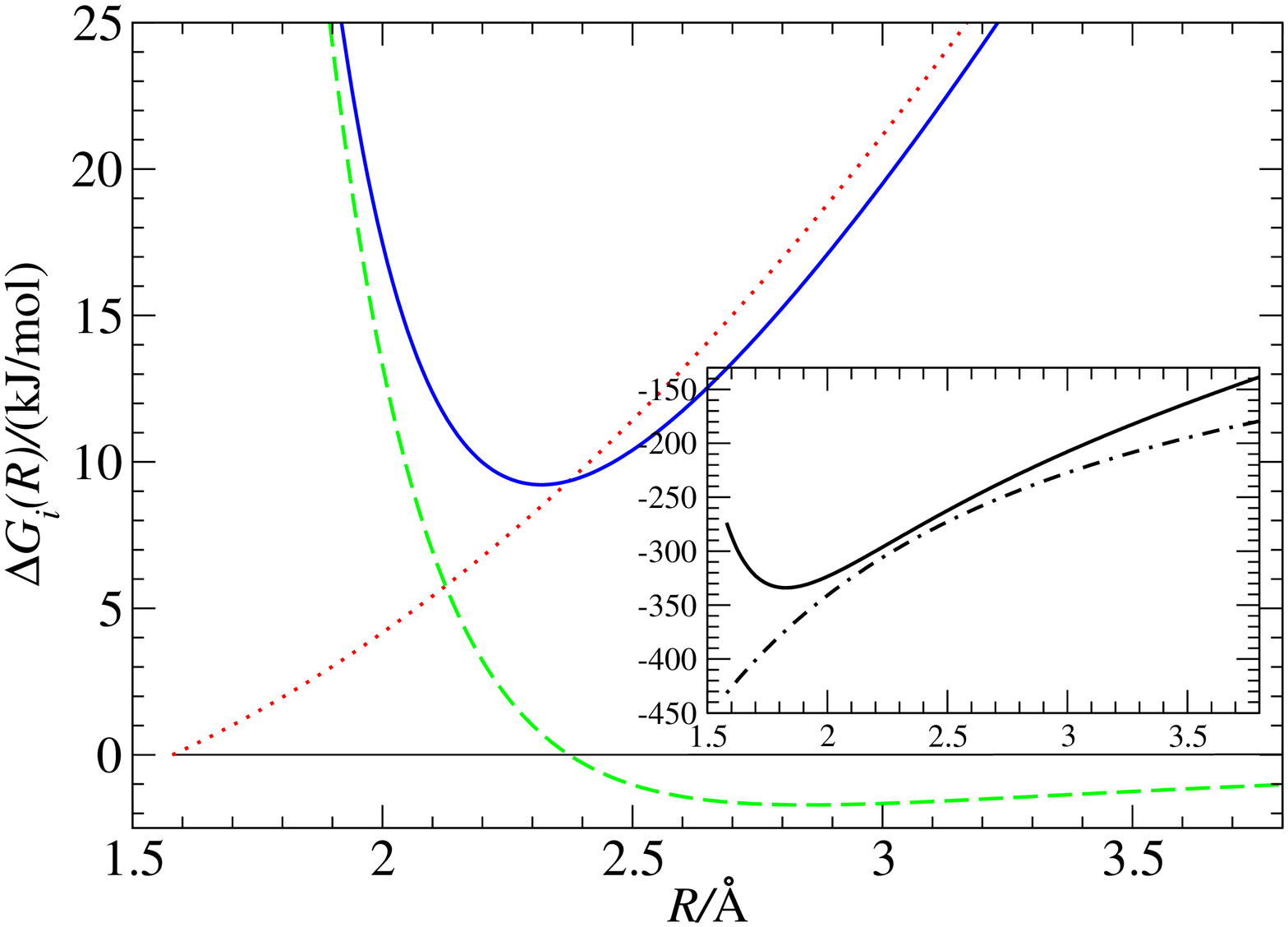, width=12cm, angle=-90}
 \end{center}
 \end{figure}

\clearpage
{\Large Figure 2}
 \begin{figure}[htb]
 \begin{center}
    \epsfig{file=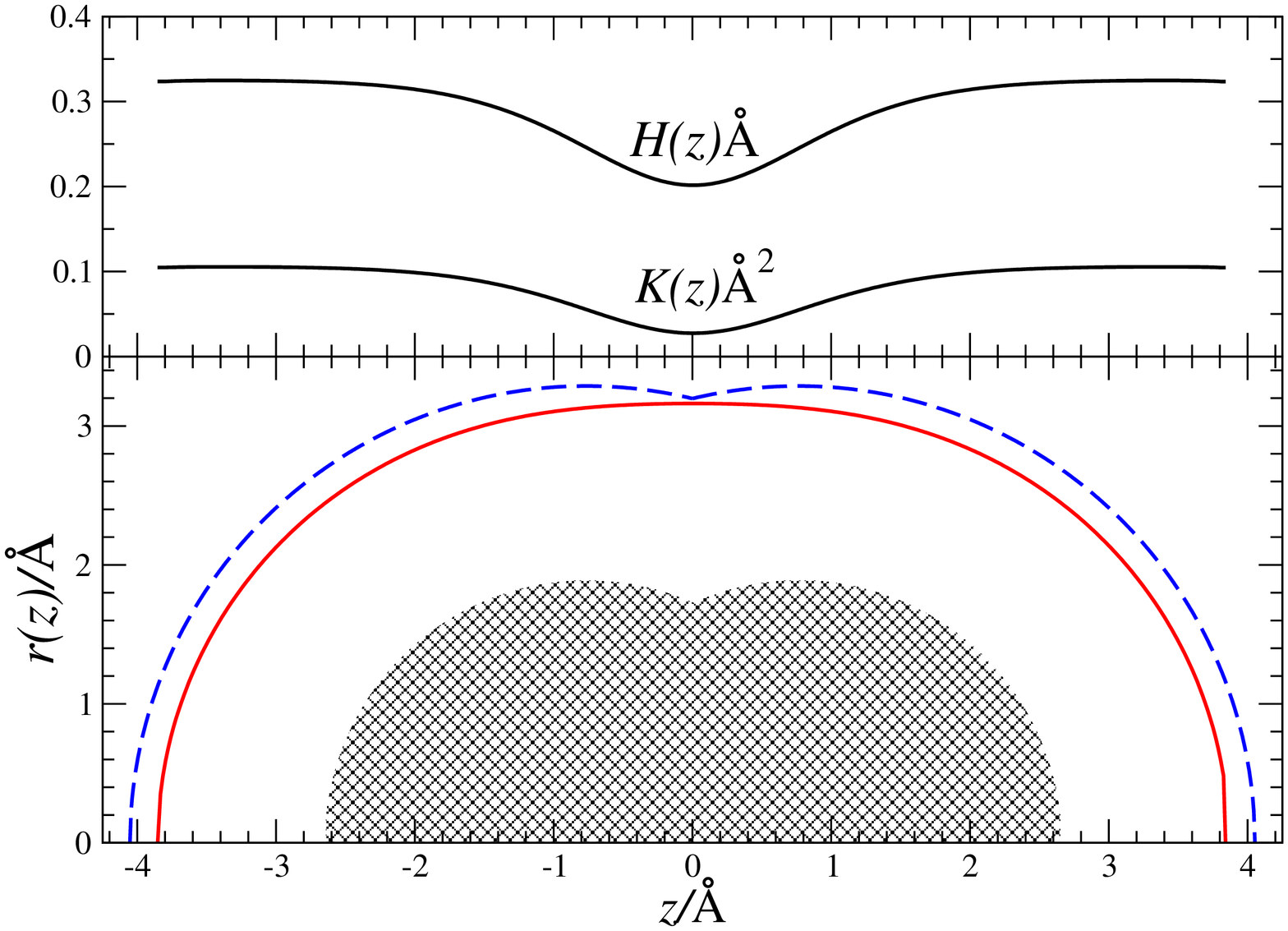, width=12cm, angle=-90}
 \end{center}
 \end{figure}

\clearpage
{\Large Figure 3}
 \begin{figure}[htb]
 \begin{center}
    \epsfig{file=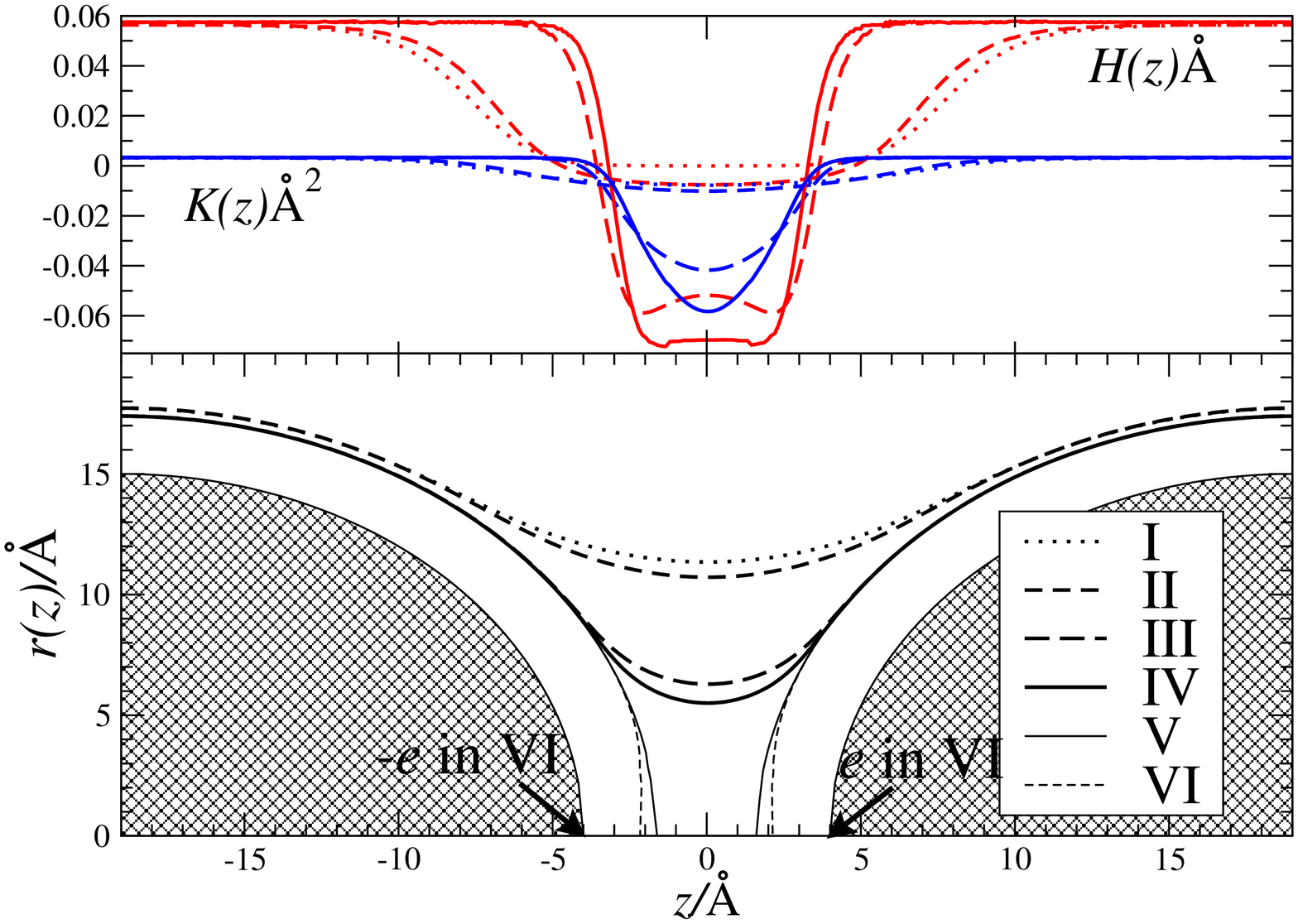, width=12cm, angle=-90}
 \end{center}
 \end{figure}

\clearpage
{\Large Figure 4}
 \begin{figure}[htb]
 \begin{center}
    \epsfig{file=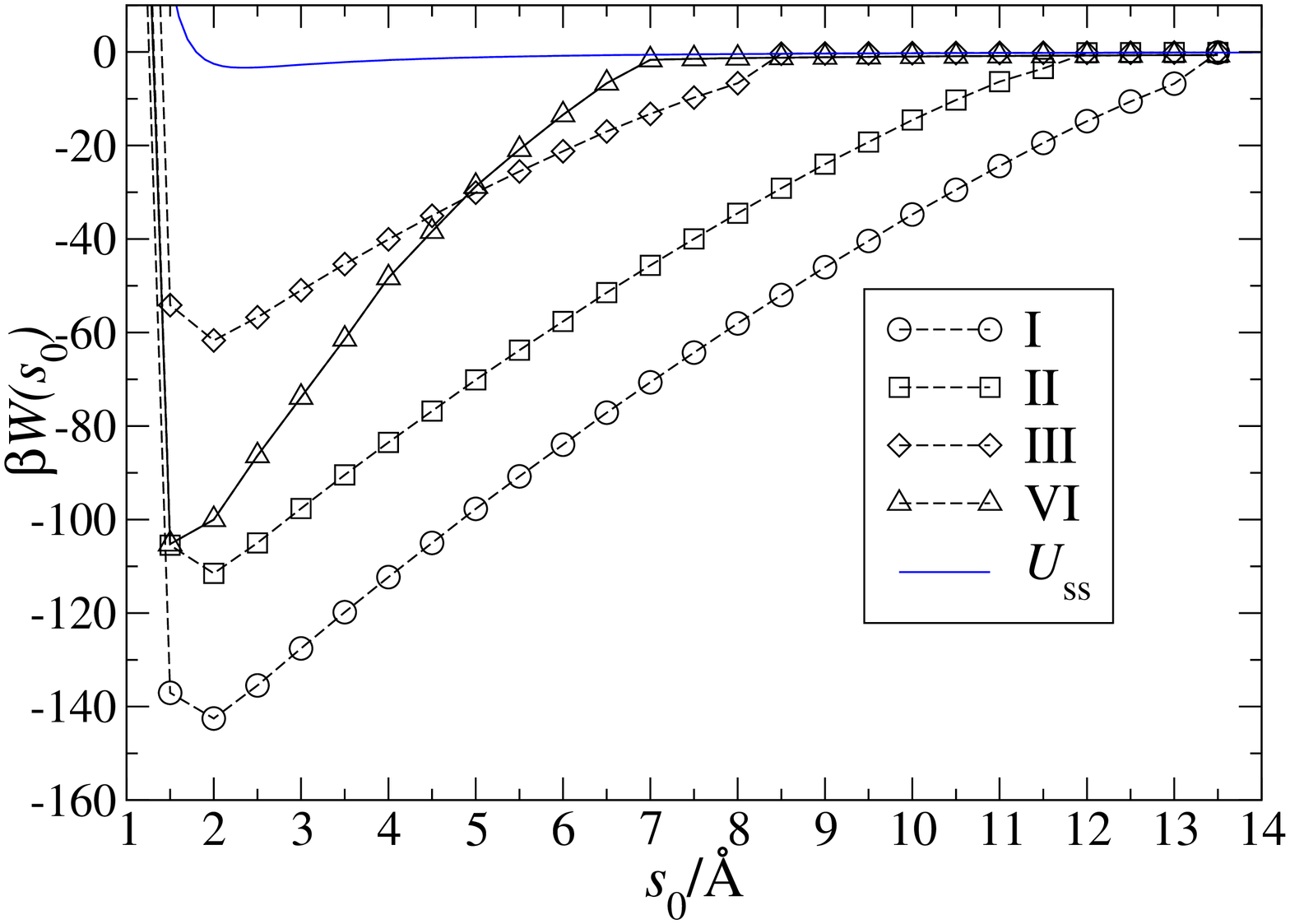, width=10cm, angle=-90} 
    \epsfig{file=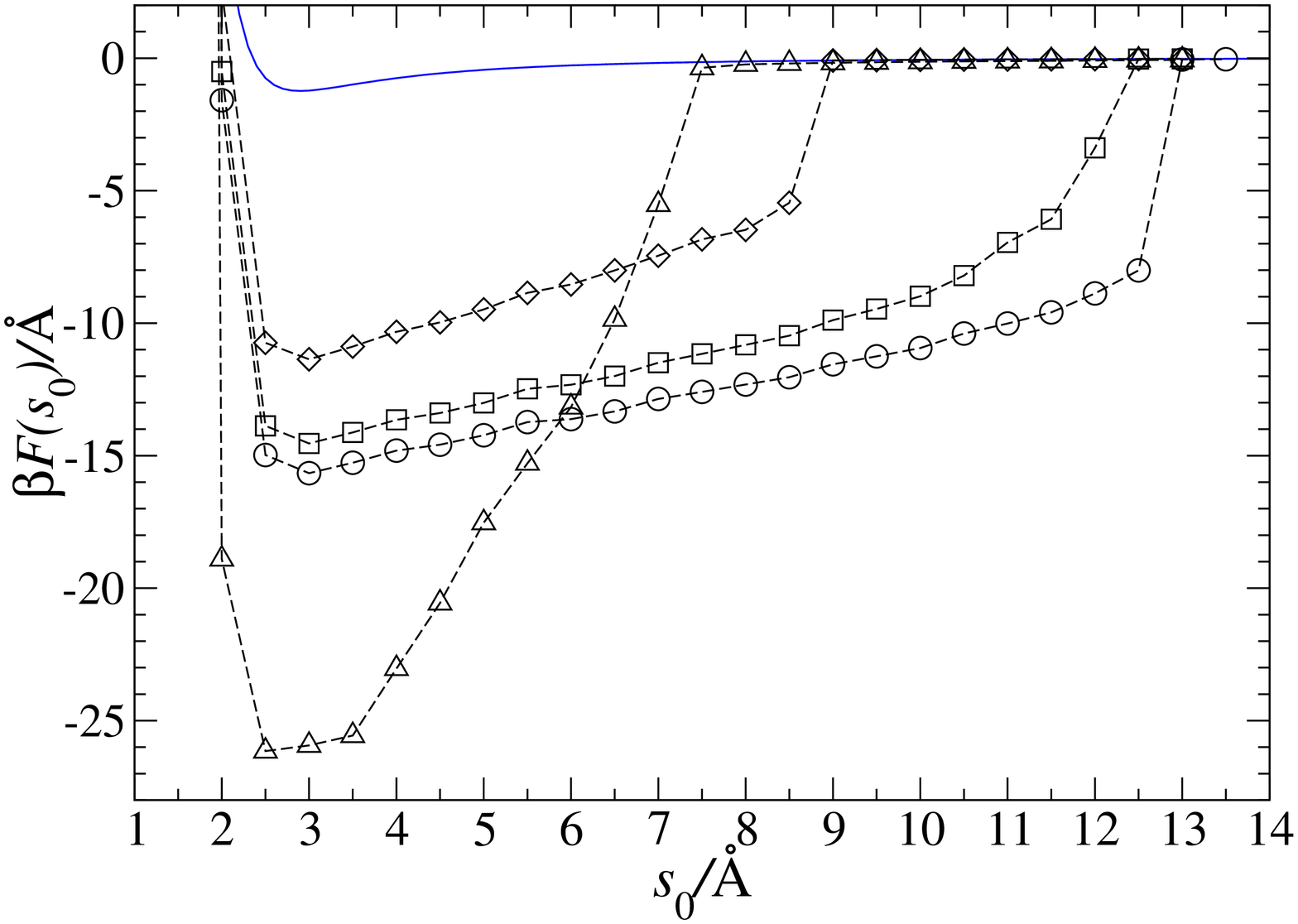,width=10cm, angle=-90}
 \end{center}
 \end{figure}

\clearpage
{\Large Figure 5}
\begin{figure}[htb]
 \begin{center}
    \epsfig{file=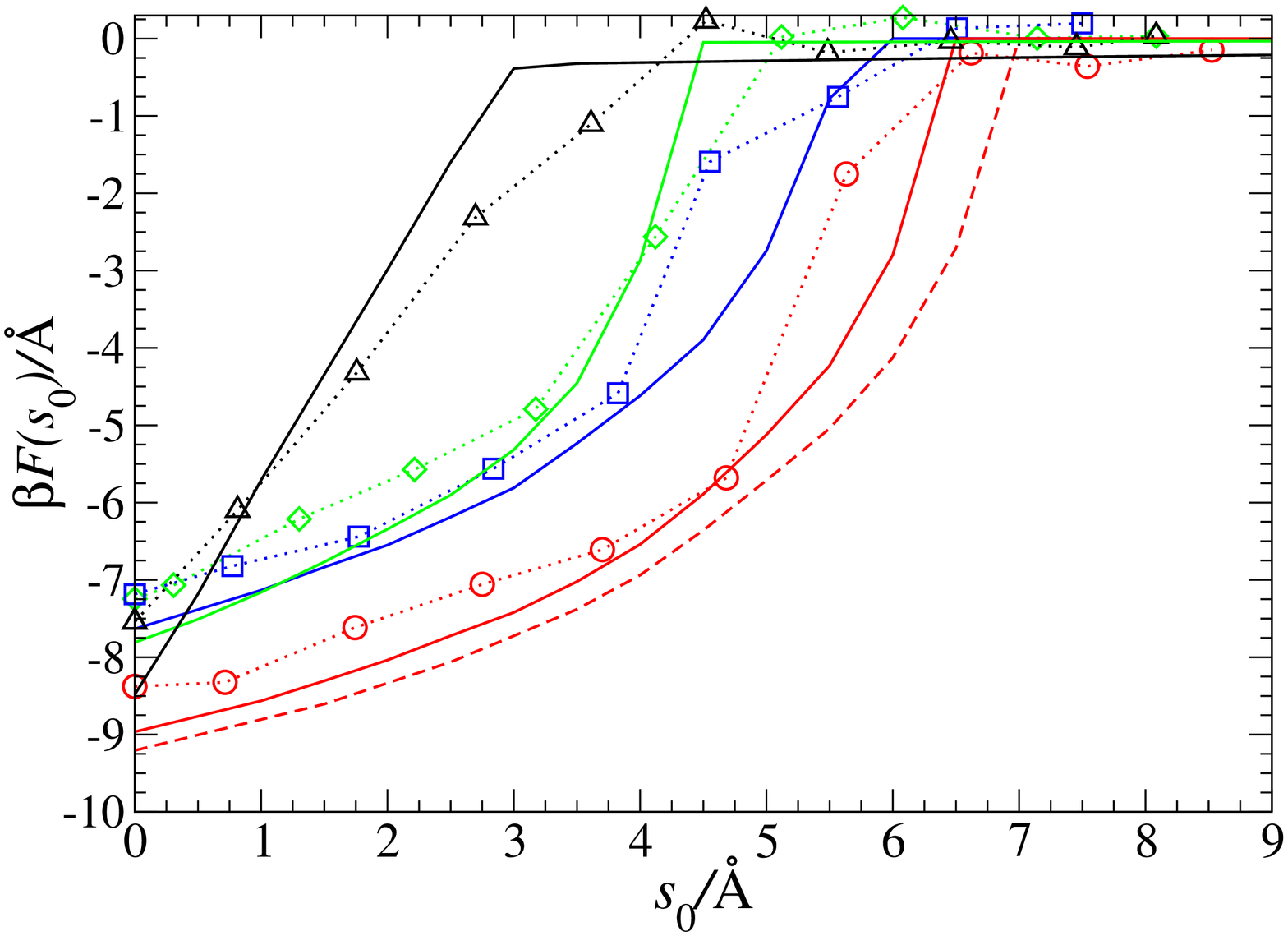,width=12cm, angle=-90}
 \end{center}
 \end{figure}

\clearpage
{\Large Figure 6}
 \begin{figure}[htb]
 \begin{center}
    \epsfig{file=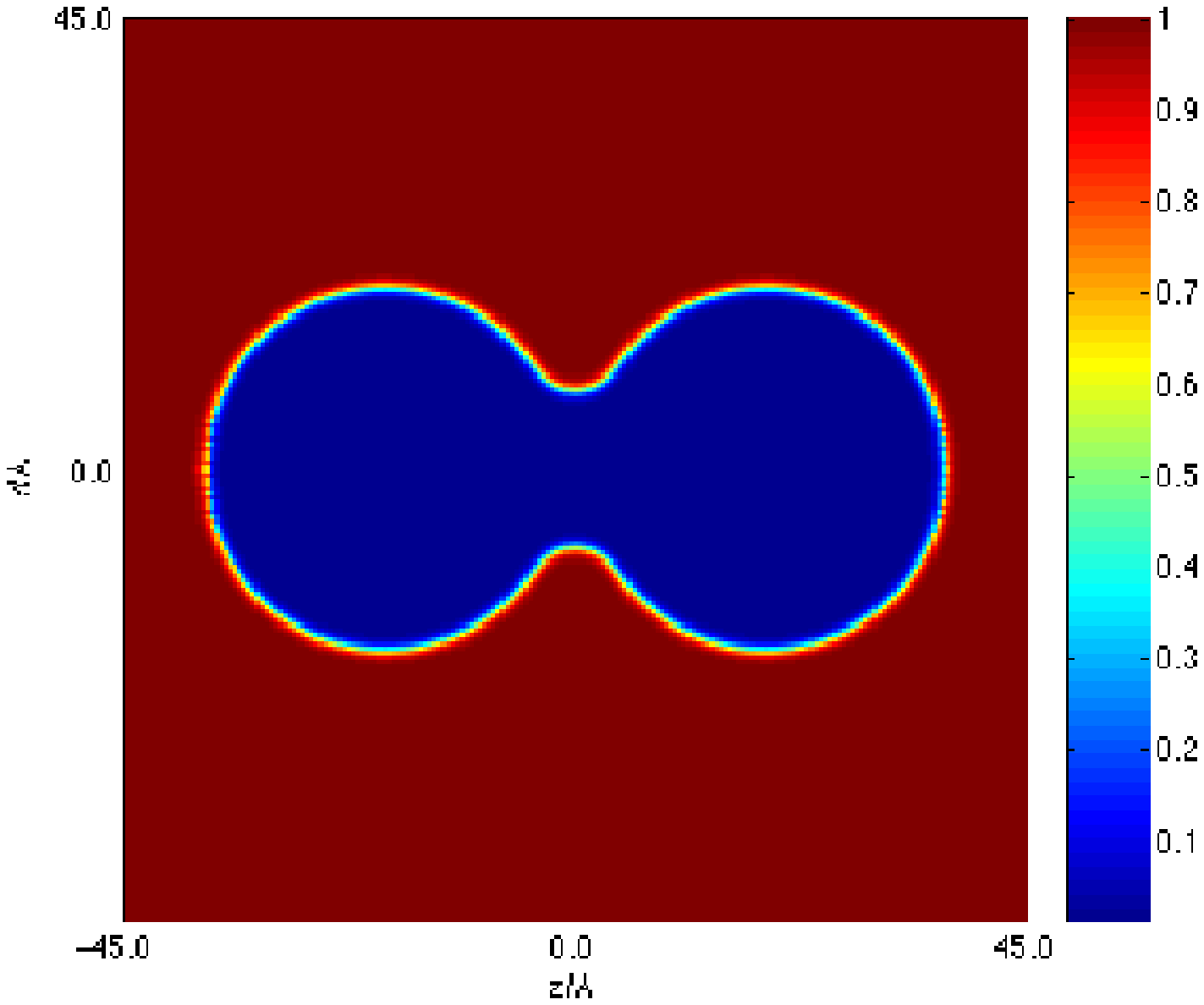, width=12cm, angle=0}
 \end{center}
 \end{figure}


\begin{thebibliography}{71}
\bibitem[{pri()}]{prince}
\bibinfo{note}{The principal curvatures are formally given by the eigenvectors
  of the Hessian (or shape operator) of $v$, $\hat S$, which can be expressed
  by the vector gradient of the unit normal vector field $\vec n(\vec r)$ of
  the surface, $\hat S=\nabla \vec n(\vec r)={\bf \nabla} \frac{\nabla v(\vec
  r)}{|\nabla v(\vec r)|}$.}
\end{thebibliography}
\end{document}